\documentclass[journal=inorgchem,manuscript=article]{achemso}

\usepackage{graphicx}
\usepackage{float}
\usepackage{color}
\usepackage{latexsym}
\title{The Crystal and Electronic Structures of Cd$_3$As$_2$, the 3D Electronic Analogue to Graphene}

\author{Mazhar N. Ali}
\email{mnali@princeton.edu}
\affiliation{Department of Chemistry, Princeton University, Princeton New Jersey 08544, USA.}
\author{Quinn Gibson}
\affiliation{Department of Chemistry, Princeton University, Princeton New Jersey 08544, USA.}
\author{Sangjun Jeon}
\affiliation{Joseph Henry Laboratories and Department of Physics, Princeton University, Princeton New Jersey 08544, USA.}
\author{Brian B. Zhou}
\affiliation{Joseph Henry Laboratories and Department of Physics, Princeton University, Princeton New Jersey 08544, USA.}
\author{Ali Yazdani}
\affiliation{Joseph Henry Laboratories and Department of Physics, Princeton University, Princeton New Jersey 08544, USA.}
\author{R. J. Cava}
\email{rcava@princeton.edu}
\affiliation{Department of Chemistry, Princeton University, Princeton New Jersey 08544, USA.}

\date{\today}

\begin{document}

\begin{abstract}

\indent{}The structure of Cd$_3$As$_2$, a high mobility semimetal reported to host electrons that act as Dirac particles, is reinvestigated by single crystal X-ray diffraction. It is found to be centrosymmetric rather than noncentrosymmetric as previously reported. It has a distorted superstructure of the antifluorite (M$_2$X) structure type with a tetragonal unit cell of \textit{a} = 12.633(3) and \textit{c} = 25.427(7) \AA~in the centrosymmetric I4$_1$/acd space group. The antifluorite superstructure can be envisioned as consisting of distorted Cd$_6$$\Box$$_2$ cubes (where $\Box$ = an empty cube vertex) in parallel columns, stacked with opposing chirality. Electronic structure calculations performed using the experimentally determined centrosymmetric structure are similar to those performed with the inversion symmetry absent, but with the important implication that Cd$_3$As$_2$ is a 3D-Dirac semimetal with no spin splitting; all bands are spin degenerate and there is a four-fold degenerate bulk Dirac point at the Fermi Energy along $\Gamma$-Z in the Brillouin zone. This makes Cd$_3$As$_2$ a 3D electronic analog to graphene. Scanning Tunneling Microscopy experiments identify a 2x2 surface reconstruction in the (112) cleavage plane of single crystals; needle crystals grow with a [110] long axis direction.  

\end{abstract}

%\pacs{} 

\maketitle

\section{Introduction}
\indent{} The topology-dependent electronic properties of solids are the subject of considerable current research. Although theoretical and experimental exploration of the electronic implications of topology fall clearly in the realm of materials physics \cite{RevModPhys.82.3045}, the compounds of interest for displaying those properties have a well-defined set of chemical characteristics, including constituent elements with similar electronegativities and strong spin orbit coupling; crystal symmetry also plays a critical role \cite{C3TC30186A}. The fact that such properties are in many cases predictable by theory has generated a wide interest in finding compounds that display them. Although the edges and surfaces of crystals and extremely thin crystals such as graphene might reasonably be expected to display electronic properties dependent on their topology \cite{König02112007,doi:10.1146/annurev-conmatphys-062910-140432,novoselov2005two}, there are also cases where topological properties have been predicted for electrons within bulk three dimensional crystals \cite{PhysRevLett.108.140405,PhysRevLett.107.186806}. Such is the case for the recently emergent compounds Na$_3$Bi and Cd$_3$As$_2$, where early characterization of the real materials indicates that this may indeed be the case \cite{borisenko2013experimental,neupane2013observation,Liu16012014}. One important prediction within this category is that in some cases the electrons will behave like they obey the ``Weyl Hamiltonian", a previously unobserved electronic state \cite{PhysRevB.83.205101,Physics.4.36}. In this instance the presence or absence of a center of symmetry in the host crystal is a critical structural characteristic for the stability of the electronic phase. 

Cd$_3$As$_2$ has been well studied in the past \cite{lin1969energy,rosenberg1959cd3as2,wallace1979electronic} but not in this context. Of particular interest have been its semimetallic character and very high electron mobility \cite{rosenberg1959cd3as2}. It is also of interest for solar cells and as an analog for graphene in exploratory device applications \cite{4154290,stoumpos2013semiconducting}. Theoretical work on the electronic structure of Cd$_3$As$_2$ has been done, but the calculations were performed using either the primitive tetragonal (P4$_2$/nmc) intermediate temperature structure (475 $^{\circ}{\rm C}$ - 600 $^{\circ}{\rm C}$)\cite{OrigCd3As2Paper} or the noncentrosymmetric low temperature (Below 475 $^{\circ}{\rm C}$) structure\cite{steigmann1968crystal}(I4$_1$cd) proposed by Steigmann and Goodyear \cite{PhysRevB.88.125427}. The primitive tetragonal structure (1935) (\textit{a} = 8.95 \AA, \textit{c} = 12.65 \AA) has As in a FCC array, with 6 Cd in fluorite-like positions - 2 of the 8 vertices of the distorted cube are fully empty. The empty vertices lie diagonally opposite each other in one face of the cube, and are ordered in a two-dimensional array. The low temperature structure (1968) has a larger unit cell (\textit{a} = 12.67 \AA, \textit{c} = 25.48 \AA) where the empty cube vertices order in a three-dimensional rather than a two-dimensional array. Recent theoretical calculations performed with the reported non-centrosymmetric low temperature I4$_1$cd structure \cite{PhysRevB.88.125427} indicate that Cd$_3$As$_2$ may be a new type of 3D-Dirac semi-metal in part due to the lack of inversion symmetry, which causes the lifting of the spin degeneracy of certain bands in the vicinity of the Dirac point, raising the possibility that Cd$_3$As$_2$ may be an example of a Weyl semimetal. Thus the presence or absence of inversion symmetry has important implications for the electronic properties of Cd$_3$As$_2$. 
\newline
\indent{}Here we re-examine the crystal structure of Cd$_3$As$_2$ using current single crystal X-ray diffraction (SXRD) and analysis methods. We also identify the growth direction of needle crystals and the cleavage plane as the [110] and (112), respectively. Through the use of Scanning Tunneling Microscopy (STM) experiments, we identify a 2x2 surface reconstruction of the (112) plane cleavage surface. We find that earlier researchers failed to appreciate the near centrosymmetricity of their reported structure, and that Cd$_3$As$_2$ in its low temperature phase is in fact centrosymmetric, with the space group I4$_1$/acd. The higher symmetry forbids any spin splitting and keeps all bands at least two fold degenerate; this has significant implications for the behavior of the electrons near the Fermi Energy. The inversion symmetry constrains Cd$_3$As$_2$ to be a non-spin-polarized 3D-Dirac semi-metal, and therefore implies that it is a 3D analog to graphene, where the electronic states are also non-spin-polarized.

\section{Experimental}
\indent{}Silvery, metallic colored single crystals of Cd$_3$As$_2$ were grown out of a Cd-rich melt with the stoichiometry (Cd$_5$)Cd$_3$As$_2$. The elements were sealed in an evacuated quartz ampoule, heated to 825 $^{\circ}{\rm C}$, and kept there for 48 hours. The sample was then cooled at a rate of 6 degrees per hour to 425 $^{\circ}{\rm C}$ and was subsequently centrifuged. Only Cd$_3$As$_2$ crystals were formed, with a crystal to flux ratio of 1:5, consistent with the equilibrium phase diagram \cite{ADMA:ADMA19910031215}. Chemical analysis of the crystals was performed in an FEI Quanta 200 FEG Environmental-SEM by energy dispersive x-ray analysis (EDX), which found them to have a Cd:As ratio of 1.500(1):1. Powder X-ray diffraction (PXRD) patterns were collected using Cu K$\alpha$ radiation on a Bruker D8 Focus diffractometer with a graphite monochromator on ground single crystals to confirm the identity of the compound as being Cd$_3$As$_2$ in the low temperature structure. The single crystal X-ray diffraction study was performed on a 0.04 mm x 0.04 mm x 0.4 mm crystal on a Bruker APEX II diffractometer using Mo K$\alpha$ radiation ($\lambda$ = 0.71073 \AA) at 100 K. Exposure time was 35 seconds with a detector distance of 60 mm. Unit cell refinement and data integration were performed with Bruker APEX2 software. A total of 1464 frames were collected over a total exposure time of 14.5 hours. 21702 diffracted peak observations were made, yielding 1264 unique observed reflections (centrosymmetric scaling) collected over a full sphere. No diffuse scattering was observed. The crystal structure was refined using the full-matrix least-squares method on F$^2$, using SHELXL2013 implemented through WinGX. An absorption correction was applied using the analytical method of De Meulenaer and Tompa \cite{deMeulenaera04926} implemented through the Bruker APEX II software. The crystal surfaces were studied with in a home-built cryogenic scanning tunneling microscope at 2 K. No twinning was observed in either the STM or single crystal diffraction measurements. Electronic structure calculations were performed in the framework of density functional theory using the Wien2k code\cite{blaha1990full} with a full-potential linearized augmented plane-wave and local orbitals basis together with the Perdew-Burke-Ernzerhof parameterization of the generalized gradient approximation\cite{perdew1996generalized}. The plane wave cutoff parameter RMTKmax was set to 7 and the Brillouin zone (BZ) was sampled by 250 k-points. The experimentally determined structure was used and spin orbit coupling (SOC) was included.

\section{Results and Discussion}
\subsection{X-ray Diffraction}
\indent{} From SXRD, a body-centered tetragonal unit cell of \textit{a} = 12.633(3) and \textit{c} = 25.427(7) was found, matching PXRD measurements. The systematic absences of the reflections in the SXRD reciprocal lattice are consistent with the possible space groups I4$_1$cd and I4$_1$/acd as previously described \cite{steigmann1968crystal}. It has previously been theorized that Cd$_3$As$_2$ has the ideal Mn$_7$SiO$_{12}$ structure type in the I4$_1$/acd space group, \cite{Zdanowicz1964} but no experimentally determined atomic positions were reported. Here, initial refinements in setting 2 of the centrosymmetric space group I4$_1$/acd were performed with all atoms on the idealized positions of the Mn$_7$SiO$_{12}$ structure type. This idealized model gave very poor fits to the data, with R1 only falling to 60$\%$. It was clear from analysis of the electron density maps that the atoms are in fact not at the ideal positions. In the final refinement, atomic positions as well as anisotropic thermal parameters for all atoms were allowed to vary freely. The refinement results are summarized in Table 1. Table 2 lists final positions for all atoms and Table 3 lists the Cd-As bond lengths. Refined atomic displacement parameters may be found in the .cif file.

%%%%%%%%%%%%%%%%%%%%%%%%%%%%%%%%%%%%%%%%%%%%%%%%%%%%%%%%%%%%%%%%%%%%%%%%%%%%%%%%%%%%%%%%%%%%%%%%%%%%%%%%%%%%%%%%%%%%%%%%%%%%%%%%%%%%%%%%%%%%%%%%%%%%%

\begin{table}[t]
\caption{\small{Refinement Parameters for Cd$_3$As$_2$}}
\scalebox{1}{
\begin{tabular}{lcccc}
\hline
\hline

Phase & Cd$_3$As$_2$ \\ 
Symmetry & Tetragonal, I4$_1$/acd (No. 142) \\ 
Cell Parameters (\AA) & \textit{a} = \textit{b} = 12.633(3), \textit{c} = 25.427(7) \\ 
  & $\alpha$ = $\beta$ = $\gamma$ = 90$^{\circ}$ \\
Wavelength (\AA) & Mo K$\alpha$ - 0.71073 \\ 
Temperature (K) & 100 \\ 
V (\AA$^3$) & 4058.0(2)  \\ 
Z & 32 \\ 
Calculated Density (g/cm$^3$)& 6.38 \\ 
Formula Weight (g/mol) & 487.1 \\ 
Absorption Coefficient (mm$^{-1}$)& 25.07 \\ 
Observations & 21702 \\ 
F000 & 6720 \\ 
Data/Restraints/Parameters & 1264/0/48 \\ 
R1 (all reflections)& 0.0480 \\ 
R1 Fo $>$ 4$\sigma$(Fo) & 0.0220 (893) \\ 
wR2 (all) & 0.040 (1264) \\ 
R$_{int}$/R($\sigma$) & 0.0617/0.0238 \\
Difference e$^-$ density (e/\AA$^3$) & 1.24/-1.25 \\ 
GooF & 1.035 \\ 

\hline
\hline
\end{tabular}}
\label{}
\end{table}

%%%%%%%%%%%%%%%%%%%%%%%%%%%%%%%%%%%%%%%%%%%%%%%%%%%%%%%%%%%%%%%%%%%%%%%%%%%%%%%%%%%%%%%%%%%%%%%%%%%%%%%%%%%%%%%%%%%%%%%%%%%%%%%%%%%%%%%%%%%%%%%%%%%%%%%%%%

\indent{}Resolving the ambiguity between a centrosymmetric crystal structure and a noncentrosymmetric crystal structure that is nearly centrosymmetric can sometimes be difficult, and is a well-known problem in crystallography\cite{MarshCentro,Ermera07176}. Unless a clear choice can be made in favor of the noncentrosymmetric model, structures must be described centrosymmetrically\cite{MarshCentro}. The authors of the previous non-centrosymmetric structure report did not appreciate the fact that an alternative origin choice would allow for a centrosymmetric structure and thus did not check a centrosymmetric model against their observed intensities, which were estimated from exposed film densities \cite{steigmann1968crystal}. In order to compare a noncentrosymmetric model for Cd$_3$As$_2$ to the centrosymmetric model, we carried out a refinement in the noncentrosymmetric I4$_1$cd space group, using the published model as a template. The final refinements (see supplementary information) with atomic positions and anisotropic displacement parameters allowed to vary yielded a wR2 of 0.0489 for 2481 data (with noncentrosymmetric scaling) with 93 parameters and an R1 = 0.0265 for 1639 Fo $>$ 4F$\sigma$. The centrosymmetric model displays significantly better R-values. The R1 value for example for all data for the I4$_1$/acd solution is 0.0481 while for the I4$_1$cd solution it is 0.0611. This is an improvement by a factor of 1.22 for the centrosymmetric structure. In addition, we used PLATON\cite{PLATON} to check for and detect missed symmetry\cite{flack2006centrosymmetric}. The ADDSYM analysis (Le Page algorithm for missing symmetry) was used on the previously reported I4$_1$cd noncentrosymmetric structure. For that noncentrosymmetric model, PLATON detected a missing inversion center at (0, $\frac{1}{4}, \frac{1}{4}$) in the unit cell and recommended a 180$^{\circ}$ rotation around the \textit{c}-axis followed by an origin shift to (0, -$\frac{1}{4}$, $\frac{1}{4}$), and a suggested space group of I4$_1$/acd. Further, the deviations of the atoms in the noncentrosymmetric model from their equivalents in the centrosymmetric model are calculated and found to be very small ($\approx$.03 \AA, supplementary info). Finally the Flack parameter (supplementary info) indicated that the structure was not noncentrosymmetric \cite{Flack:a22047}. This analysis thus further confirmed that the centrosymmetric crystal structure is the correct one.

%%%%%%%%%%%%%%%%%%%%%%%%%%%%%%%%%%%%%%%%%%%%%%%%%%%%%%%%%%%%%%%%%%%%%%%%%%%%%%%%%%%%%%%%%%%%%%%%%%%%%%%%%%%%%%%%%%%%%%%%%%%%%%%%%%%%%%%%%

\subsection{Crystal Structure}

\indent{}Cd$_3$As$_2$ has different but related crystal structures as a function of temperature, which all can be considered defect antifluorite types; the Cd is distributed in the cube-shaped array occupied by F in CaF$_2$, while the As is in the FCC positions that are occupied by Ca. Thus in Cd$_3$As$_2$, the formally Cd$^{2+}$ ions are four-coordinated by As and the formally As$^{3-}$ ions are eight-coordinated by Cd. With As$^{3-}$ in VI fold coordination expected to have a radius of about 2.22 \AA~and Cd$^{2+}$ in IV coordination expected to have a radius of about 0.92 \AA, the $\frac{r_{Cd^{2+}}}{r_{As^{3-}}} = 0.41$, which is near the ideal 0.414 cutoff for tetrahedral coordination of the metal in the CaF$_2$-type structure ($\frac{r_M}{r_X}$ = 0.15 - 0.414).\cite{muller1993inorganic}. Further, as evidenced by the stoichiometry, Cd$_3$As$_2$ is Cd deficient of the ideal Cd$_4$As$_2$ antifluorite formula, missing $\frac{1}{4}$ of the Cd atoms needed to form a simple cube around the As. Thus the (distorted) cube can be written as Cd$_6$$\Box$$_2$, where $\Box$ = an empty vertex. The As and Cd coordinations in our Cd$_3$As$_2$ structure are shown in Figure 1. In the ordered, lower temperature variants, the Cd atoms shift from the ideal antifluorite positions toward the empty vertices of the cube (shown in Figure 2a). This distortion makes occupancy of the empty vertices highly energetically unfavored. At high temperatures (T $>$ 600 $^{\circ}{\rm C}$) the Cd ions are disordered\cite{alphaCd3As2} and so Cd$_3$As$_2$ adopts the ideal antifluorite space group Fm-3m with \textit{a} = 6.24 \AA~(The HT structure). Between 600 $^{\circ}{\rm C}$ and 475 $^{\circ}{\rm C}$, the intermediate temperature (IT) P4$_2$/nmc structure is found, where the Cd ions order so that the empty vertices are located on diagonally opposite corners of one face of the Cd$_6$$\Box$$_2$ cubes (see below). These cubes then stack so that the empty vertices form channels parallel to the \textit{a} and \textit{b} axes at different levels along the \textit{c} axis\cite{alphaCd3As2}. On further cooling, another ordering scheme is found. Below 475 $^{\circ}{\rm C}$, the Cd atoms further order in a three-dimensional fashion, such that each distorted Cd$_6$$\Box$$_2$ cube stacks on top of the previous one after a 90$^{\circ}$ rotation (either clockwise or counter-clockwise depending on the particular chain) about the stacking axis (parallel to the \textit{c} axis). This is the low temperature (LT) centrosymmetric I4$_1$/acd structure of the crystals whose structure is determined here and whose physical properties are of current interest. This also appears to be a new structure type. Figures 2b and 2c show how the three structures are related: the P4$_2$/nmc structure is a supercell of the disordered Fm-3m structure, and the I4$_1$/acd structure is a supercell of the P4$_2$/nmc structure. 

%%%%%%%%%%%%%%%%%%%%%%%%%%%%%%%%%%%%%%%%%%%%%%%%%%%%%%%%%%%%%%%%%%%%%%%%%%%%%%%%%%%%%%%%%%%%%%%%%%%%%%%%%%%%%%%%%%%%%%%%%%%%%%%%%%%%%%%%%%%%%%%%%%%%%%%%%%%%%%%%%%%%%%%%

\begin{table}[t]
\caption{\small{Refined atomic positions for Cd$_3$As$_2$ in space group I4$_1$/acd}}
\scalebox{1}{
\begin{tabular}{ccccc}
\hline
\hline
Atom & Wyckoff & \textit{x} & \textit{y} & \textit{z} \\[1ex]
\hline 

Cd1 & 32g & 0.13955(3) & 0.36951(3) & 0.05246(2)  \\ 
Cd2 & 32g & 0.11162(3) & 0.64230(3) & 0.07243(2)  \\ 
Cd3 & 32g & 0.11863(3) & 0.10598(4) & 0.06247(2)  \\ 
As1 & 32g & 0.24602(4) & 0.25779(5) & 0.12315(2)  \\ 
As2 & 16d & 0 & $\frac{1}{4}$ & 0.99931(2)  \\ 
As3 & 16e & $\frac{1}{4}$ & 0.51070(7) & 0 \\ 

\hline
\hline

\end{tabular}}

\label{}
\end{table}

%%%%%%%%%%%%%%%%%%%%%%%%%%%%%%%%%%%%%%%%%%%%%%%%%%%%%%%%%%%%%%%%%%%%%%%%%%%%%%%%%%%%%%%%%%%%%%%%%%%%%%%%%%%%%%%%%%%%%%%%%%%%%%%%%%%%%%%%%%%%%%%%%%%%%%%%%%

\begin{table}[h]
\caption{\small{Bond lengths in Cd$_3$As$_2$}}
\scalebox{1}{
\begin{tabular}{ccc|cccc}
\hline
\hline
Atom1 & Atom2 & Distance \AA & & Atom1 & Atom2 & Distance \AA \\[1ex]
\hline 

As1 & Cd3 & 2.6250(8) & & Cd1 & As3 & 2.6282(8)  \\ 
& Cd2 & 2.6507(8) & & & As1 & 2.6518(8) \\
& Cd1 & 2.6518(8) & & & As2 & 2.6858(7) \\
& Cd2 & 2.6552(8) & & & As1 & 2.9838(8)  \\
& Cd3 & 2.9408(8) \\
& Cd1 & 2.9837(8) \\
& & \\
As2 & Cd2 & 2.6771(8) & & Cd2 & As1 & 2.6507(8)  \\ 
& Cd2 & 2.6771(8) & & & As1 & 2.6552(8) \\
& Cd1 & 2.6858(7) & & & As2 & 2.6771(8) \\
& Cd1 & 2.6858(7) & & & As3 & 3.0351(8) \\
& Cd3 & 2.8522(7) \\
& Cd3 & 2.8522(7) \\
& & \\
As3 & Cd3 & 2.5935(7) & & Cd3 & As3 & 2.5935(7)   \\ 
& Cd3 & 2.5935(7) & & & As1 & 2.6250(8) \\
& Cd1 & 2.6282(8) & & & As2 & 2.8522(7) \\
& Cd1 & 2.6282(8) & & & As1 & 2.9408(8) \\
& Cd2 & 3.0351(8) \\
& Cd2 & 3.0351(8) \\

\hline
\hline

\end{tabular}}

\label{}
\end{table}

%%%%%%%%%%%%%%%%%%%%%%%%%%%%%%%%%%%%%%%%%%%%%%%%%%%%%%%%%%%%%%%%%%%%%%%%%%%%%%%%%%%%%%%%%%%%%%%%%%%%%%%%%%%%%%%%%%%%%%%%%%%%%%%%%%%%%%%%%%%%%%%%%%%%%%%%%%

\indent{}The I4$_1$/acd structure has 3 unique Cd atoms and 3 unique As atoms and is schematically shown in Figure 3. Since the empty vertices of the Cd$_6$$\Box$$_2$ cubes sit diagonally opposite each other in a face of the distorted cube, the incomplete Cd cube can be thought of as having only one ``closed face". This ``closed face" changes position in either a clockwise or counter-clockwise fashion as the cubes stack along the \textit{c} direction, resulting in a screwing chain of cubes. These chains then align so that each chain is next to a chain of an opposite handedness; a right-handed chain is surrounded by left-handed chains. This results in an inversion center being present between each chain.

%%%%%%%%%%%%%%%%%%%%%%%%%%%%%%%%%%%%%%%%%%%%%%%%%%%%%%%%%%%%%%%%%%%%%%%%%%%%%%%%%%%%%%%%%%%%%%%%%%%%%%%%%%%%%%%%%%%%%%%%%%%%%%%%%%%%%%%%%%%%%%%%%%%%%%%%%%

\indent{}For physical property measurements, it is important to identify characteristic planes and directions in the as-grown crystals. Cd$_3$As$_2$ grown as described here forms both irregular and rod-like crystals, all of which appear to have planar pseudo-hexagonal surfaces, often in steps, perpendicular to the long crystal axis. While the detailed SXRD measurements for structural refinement were carried out on small crystals, which gave the cleanest diffraction spots, larger crystals are employed for property measurements. Several of these larger crystals (e.g. with typical dimensions of 0.15 mm by 0.05 mm by 1.2 mm) were used in order to ascertain the growth direction of the needles. The crystals were mounted onto flat kapton holders and the Bruker APEX II software\cite{APEXII} was used to indicate the face normals of the crystal after the unit cell and orientation matrix were determined. The largest crystals were also placed onto PXRD slides and oriented diffraction experiments were conducted. The long axis of the needles was consistently found to be the [110] direction. The planar pseudo-hexagonal crystal surfaces with normals perpendicular to the growth direction are the (112) planes; these correspond to the close packed planes in the cubic antifluorite phase transformed into the LT structure supercell Miller indices. STM studies find that the cleavage plane of Cd$_3$As$_2$ is the pseudo-hexagonal (112) plane found here as a well-developed face in the bulk crystals. Figure 4a shows a projection of the structure with the (112) plane shaded in orange. Figure 4b is a topographic STM image (V$_{bias} =$ -250 mV, I = 50 pA, Temp = 2 K) of the (112) cleavage plane. The nearest neighbor spacing is found to 4.4$\pm$0.1 \AA, which is consistent with the As-As spacing or Cd-Cd spacing on a (112) type plane. Also visible in the inset of Figure 4b is the appearance of a 2x2 surface reconstruction, likely due to dangling bonds from the termination of the (112) plane during cleavage. Figure 4c is a projection slightly off of the [110] direction, with the corresponding lattice plane shaded in green. Figure 4d shows the unit cell axes and the growth directions of a large needle crystal of Cd$_3$As$_2$ that is suitable for physical property measurements.

\indent{}A [110] needle direction for a tetragonal symmetry crystal is relatively uncommon \cite{buerger1956elementary}. The crystal growth conditions employed in this study were such that the majority of the slow cooling took place through the stable temperature region of the IT P4$_2$/nmc structure. We hypothesize that the fastest growing direction in this phase is along the chain axis of the empty cube vertices, which is a principal axis of the structure (either the a or b axis). Thus the needle crystals have already grown before the last 50 degrees of the cooling, during which the low temperature annealing results in the ordering of the Cd atoms that yields the I4$_1$/acd structure.  As can be seen from Figure 2, the principal axes of the P4$_2$/nmc structure become the set of [110] directions in the I4$_1$/acd structure. Variation of the growth temperature and conditions in future crystal growth studies may result in different types of needle axes, but the (112) close packed plane for crystal face development and cleavage is likely to be strongly preferred.

%%%%%%%%%%%%%%%%%%%%%%%%%%%%%%%%%%%%%%%%%%%%%%%%%%%%%%%%%%%%%%%%%%%%%%%%%%%%%%%%%%%%%%%%%%%%%%%%%%%%%%%%%%%%%%%%%%%%%%%%%%%%%%%%%%%%%%%%%%%%%%%%%%%%%%%%%%

\subsection{Electronic Structure}

\indent{}The electronic structure calculated from the experimentally determined centrosymmetric structure found here (Table 2) is shown in Figure 5. This band structure is similar to that reported previously for the incorrect noncentrosymmetric crystal structure\cite{PhysRevB.88.125427}. Crucially, however, due to the inversion symmetry present in the I4$_1$/acd structure, there is no spin splitting around the Dirac point (where the bands cross between $\Gamma$ and Z at E$_F$ and all bands are at least two-fold degenerate. Thus, the centrosymmetric structure indicates that Cd$_3$As$_2$ is a 3D-Dirac semi-metal with two fold degenerate bands that come together to a four-fold degenerate Dirac point at the Fermi level. It is therefore the 3D analog to graphene, where there is also no spin-splitting at the Dirac point. Furthermore, the states at the $\Gamma$ point can now be characterized by their full symmetry (including their inversion eigenvalue) thus allowing parity counting to demonstrate the nontrivial topology in this ground state. 

%%%%%%%%%%%%%%%%%%%%%%%%%%%%%%%%%%%%%%%%%%%%%%%%%%%%%%%%%%%%%%%%%%%%%%%%%%%%%%%%%%%%%%%%%%%%%%%%%%%%%%%%%%%%%%%%%%%%%%%%%%%%%%%%%%%%%%%%%%%%%%%%%%%%%%%%%%
%CONCLUSION%
\section{Conclusion}
\indent{}In conclusion, we report the correct crystal structure of Cd$_3$As$_2$ in the low temperature, three-dimensionally ordered phase, as well as the corresponding electronic structure. We identify that for crystals grown as reported here, the needle growth direction is the [110] and the cleavage plane and most developed face in crystals is the pseudo-hexagonal (112) plane. Also present is a 2x2 surface reconstruction of the (112) plane cleavage surface. We found that Cd$_3$As$_2$ crystallizes in the centrosymmetric, I4$_1$/acd space group, and as such appears to be a new structure type. The Cd$_6$$\Box$$_2$ cubes order in a spiral, corkscrew fashion along an axis parallel to \textit{c}. Each corkscrew is surrounded by corkscrews of the opposite handedness, which results in the overall structure having inversion symmetry. The previously reported model in the I4$_1$cd space group (\# 110), where an inversion center was omitted, can be related to this structure by rotating by 180$^{\circ}$ about the c-axis and then shifting the origin to 0, -$\frac{1}{4}$, $\frac{1}{4}$. Since the correct centrosymmetric structural model uses only 6 unique atoms to describe the structure, electronic structure calculations become much more facile, which will help with the theoretical analysis of the electronic structure of this phase. In electronic structure calculations based on the centrosymmetric crystal structure, we find that the previously reported bulk band crossing along $\Gamma$-Z at the Fermi level is maintained, however due to the inversion symmetry, no spin splitting is allowed. Therefore Cd$_3$As$_2$ is expected to be a non-spin-split 3D-Dirac semi-metal, and a three-dimensional analog to graphene.

\bigskip 
\begin{acknowledgement}

This research was supported by the Army Research Office, grant W911NF-12-1-0461.

\end{acknowledgement}

%%%%%%%%%%%%%%%%%%%%%%%%%%%%%%%%%%%%%%%%%%%%%%%%%%%%%%%%%%%%%%%%%%%%%%
\bibliography{Lit}
\newpage

%FIGURE 1%%%%%%%%%%%%%%%%%%%%%%%%%%%%%%%%%%%%%%%%%%%%%%%%%%%%%%%%%%%%%%%%%%%%%%%%%%%%%%%%%%%%%%%%%%%%%%%%%%%%%%%%%%%%%%%%%%%%%%%%%%%%%%%%%%%%%%%%%%%%%%

\begin{figure}[h]
	\includegraphics[width=0.45\textwidth]{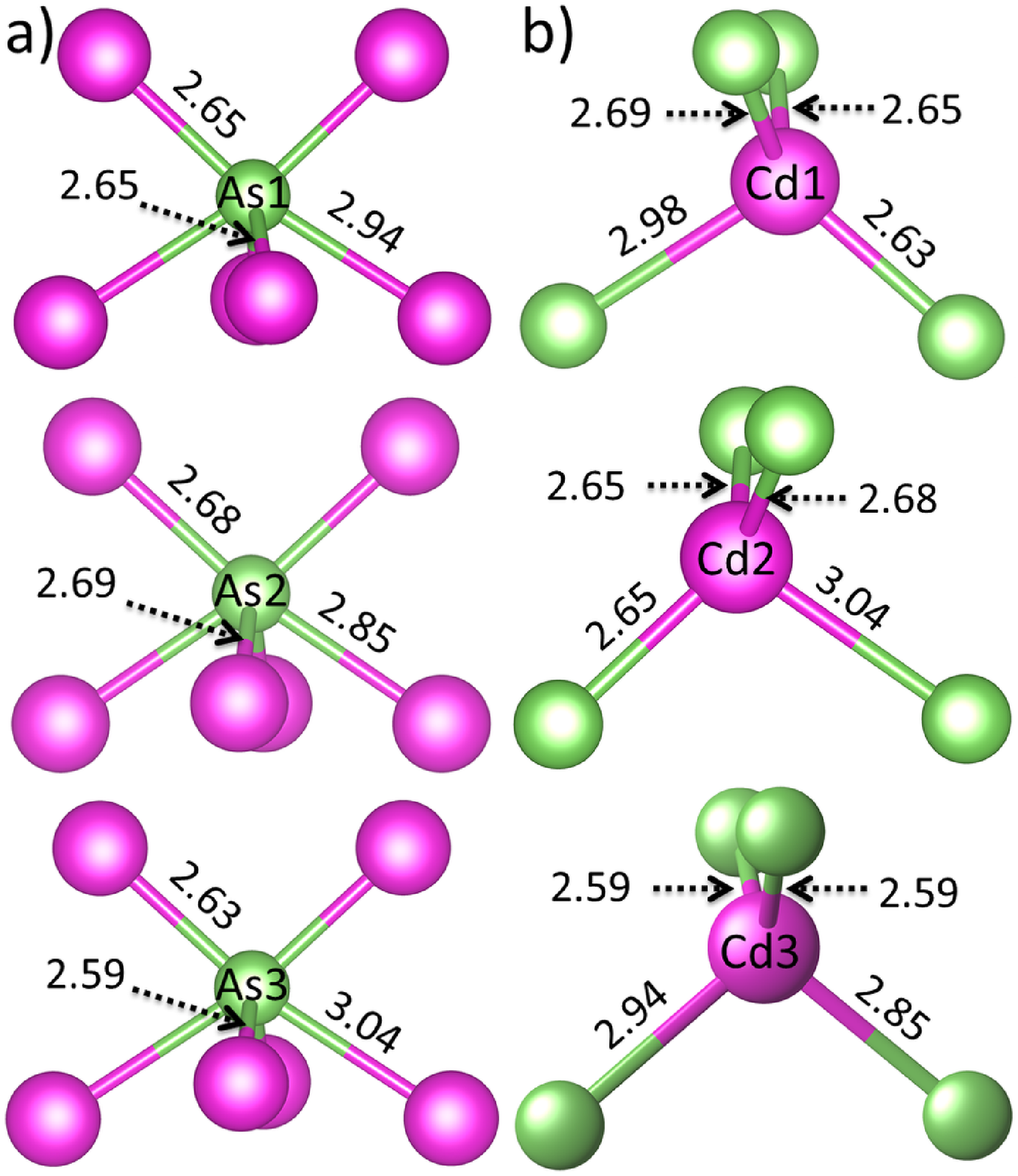}
	\caption{\scriptsize{\textbf{(color online):} a) The three coordination environments around the As atoms in Cd$_3$As$_2$. Due to the two vacancies per As-Cd cube, the remaining Cd atoms shift toward the vacancies resulting in different bond lengths. b) The coordination of As around the Cd atoms, where the As form regular tetrahedra but the Cd atoms sit off-center.}}
	\label{Figure_1}
\end{figure}
\newpage

%FIGURE 2%%%%%%%%%%%%%%%%%%%%%%%%%%%%%%%%%%%%%%%%%%%%%%%%%%%%%%%%%%%%%%%%%%%%%%%%%%%%%%%%%%%%%%%%%%%%%%%%%%%%%%%%%%%%%%%%%%%%%%%%%%%%%%%%%%%%%%%%%%%%%%

\begin{figure}[h]
	\includegraphics[width=0.9\textwidth]{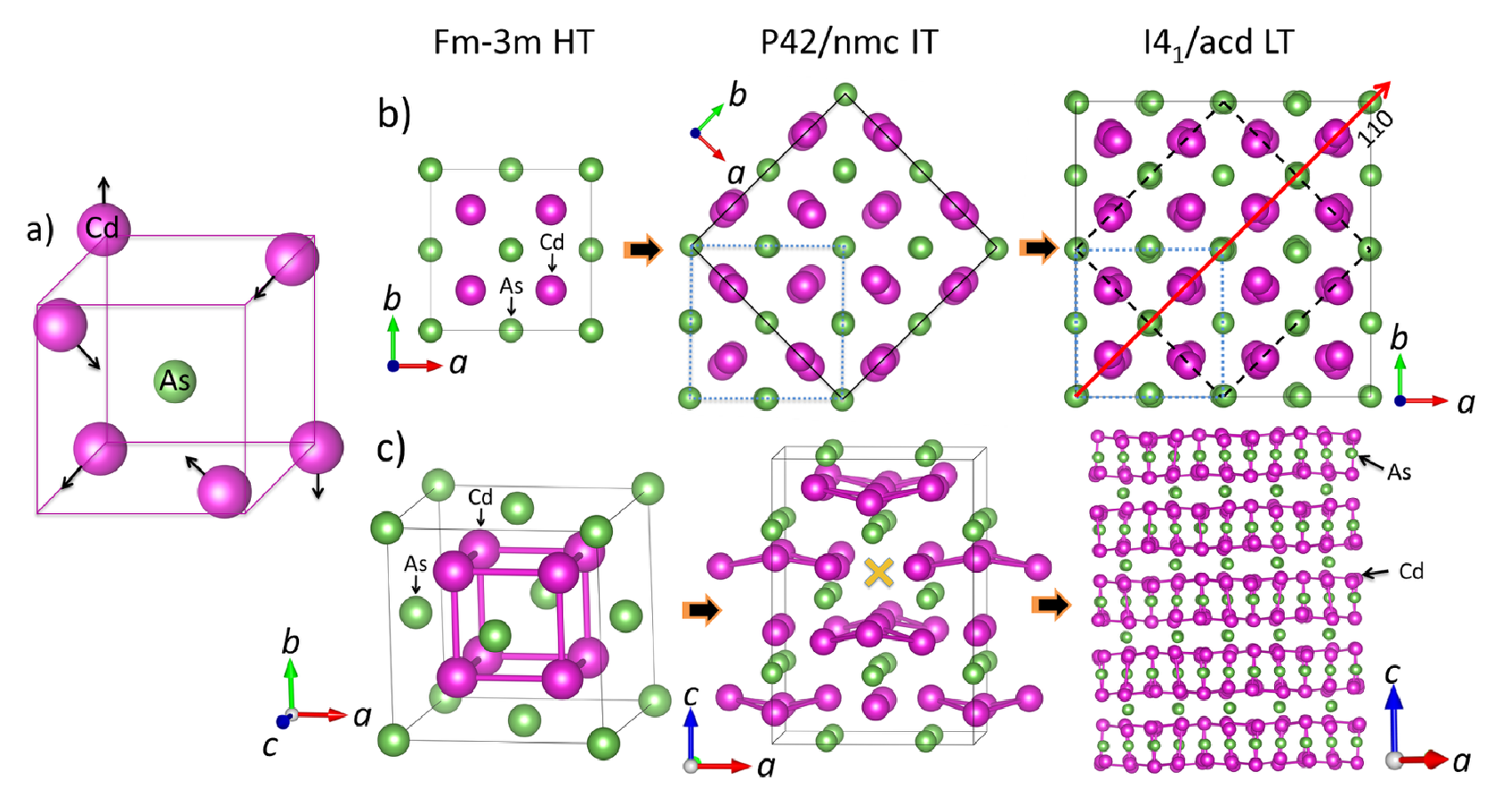}
	\caption{\scriptsize{\textbf{(color online):} a) Single As-Cd cube with the Cd atoms displaced towards the Cd vacancies, resulting in distorted cubes. These cubes stack in corkscrews along the \textit{c}-axis in the LT form. b) The relationship of the cubic Cd$_3$As$_2$ structure to the ordered lower temperature derivatives. Left to right, the high temperature (HT) Fm-3m structure, the intermediate temperature (IT) structure (P4$_2$/nmc), followed by the centrosymmetric I4$_1$/acd low temperature (LT) structure. c) The Cd cubes and their ordering for each structure. In the cubic phase, the Cd atoms are disordered over all of the pseudo-cube sites. In the P4$_2$/nmc structure the Cd vacancy channels are shown running along the \textit{b} axis with the x symbol. The I4$_1$/acd structure is shown along the [110] direction, which is equivalent to the channel direction in the P4$_2$/nmc structure, to illustrate that chains of vacancies are no longer present.}}
	\label{Figure_2}
\end{figure}
\newpage

%FIGURE 3%%%%%%%%%%%%%%%%%%%%%%%%%%%%%%%%%%%%%%%%%%%%%%%%%%%%%%%%%%%%%%%%%%%%%%%%%%%%%%%%%%%%%%%%%%%%%%%%%%%%%%%%%%%%%%%%%%%%%%%%%%%%%%%%%%%%%%%%%%%%%%%%%

\begin{figure}[h]
	\includegraphics[width=0.65\textwidth]{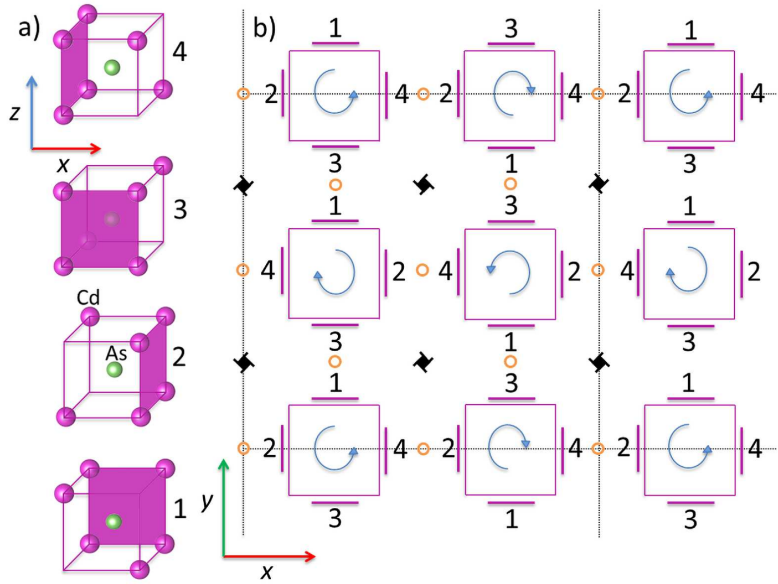}
	\caption{\scriptsize{\textbf{(color online):} a) As-Cd cubes with Cd vacancies diagonally opposite each other in one face of each cube, resulting in a single closed face of each cube (shaded). Stacking along c-axis is shown with the numbers corresponding to the layer in which each cube sits. The pink and green spheres represent Cd and As, respectively. b) Schematic diagram of the corkscrew stacking of As-Cd cubes; \textit{c}-axis is coming out of the page. The squares are the projection of the cubes looking down the \textit{c}-axis and the bolded lines with the numbers beside them represent the closed face of the cube and the layer in which it sits. Four layers of As-Cd cubes stack before the pattern repeats. Orange circles show the inversion centers, black spiral symbols show the screw axis and the dotted black line is the unit cell.}}
	\label{Figure_3}
\end{figure}
\newpage

%FIGURE 4%%%%%%%%%%%%%%%%%%%%%%%%%%%%%%%%%%%%%%%%%%%%%%%%%%%%%%%%%%%%%%%%%%%%%%%%%%%%%%%%%%%%%%%%%%%%%%%%%%%%%%%%%%%%%%%%%%%%%%%%%%%%%%%%%%%%%%%%%%%%%%%%%%%

\begin{figure}[h]
	\includegraphics[width=0.85\textwidth]{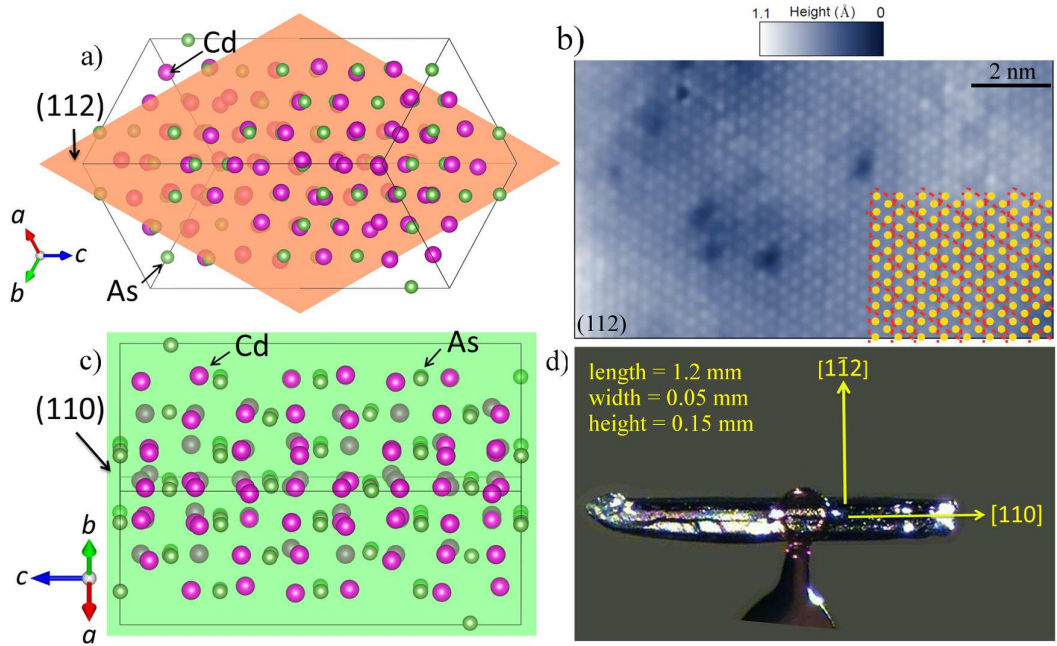}
	\caption{\scriptsize{\textbf{(color online):} a) Projection of Cd$_3$As$_2$ into the pseudo-hexagonal (112) plane, which is shaded in orange. This is the cleavage plane, and also a developed large facet in bulk crystals. b) Topographic STM image of the cleaved surface of a Cd$_3$As$_2$ crystal. The pseudo-hexagonal symmetry corresponding to the (112) plane is seen. The inset shows (orange dots) the (112) plane projection with nearest neighbor spacing of 4.4$\pm$0.1 \AA~and the red dashed lines identify a 2x2 surface reconstruction. c) Projection along the [110] direction of Cd$_3$As$_2$ with the plane normal shaded in green. This is the needle axis direction. d) A single crystal of Cd$_3$As$_2$ with the needle growth direction and "plate" normal direction labeled ([110] and [1-12], respectively). }}
	\label{Figure_4}
\end{figure} 
\newpage

%FIGURE 5%%%%%%%%%%%%%%%%%%%%%%%%%%%%%%%%%%%%%%%%%%%%%%%%%%%%%%%%%%%%%%%%%%%%%%%%%%%%%%%%%%%%%%%%%%%%%%%%%%%%%%%%%%%%%%%%%%%%%%%%%%%%%%%%%%%%%%%%%%%%%%%%%%%%%%%%%%%%%%%%%%

\begin{figure}[h]
	\includegraphics[width=0.5\textwidth]{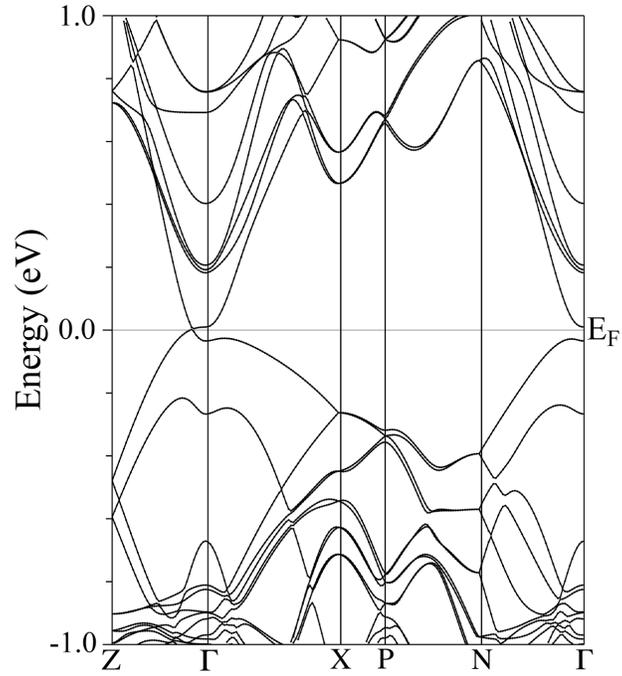}
	\caption{\scriptsize{\textbf{(color online):} The calculated electronic structure of Cd$_3$As$_2$. A symmetry allowed crossing is observed along $\Gamma$-Z at E$_F$, representative of the Dirac point. No spin splitting is allowed due to the presence of inversion symmetry.}}
	\label{Figure_5}
\end{figure}
\newpage

%%%%%%%%%%%%%%%%%%%%%%%%%%%%%%%%%%%%%%%%%%%%%%%%%%%%%%%%%%%%%%%%%%%%%%%%%%%%%%%%%%%%%%%%%%%%%%%%%%%%%%%%%%%%%%%%%%%%%%%%%%%%%%%%%%%%%%%%%%%%%%%%%%%%%%%%%%
Synopsis

\begin{figure}[h]
	\includegraphics[width=0.9\textwidth]{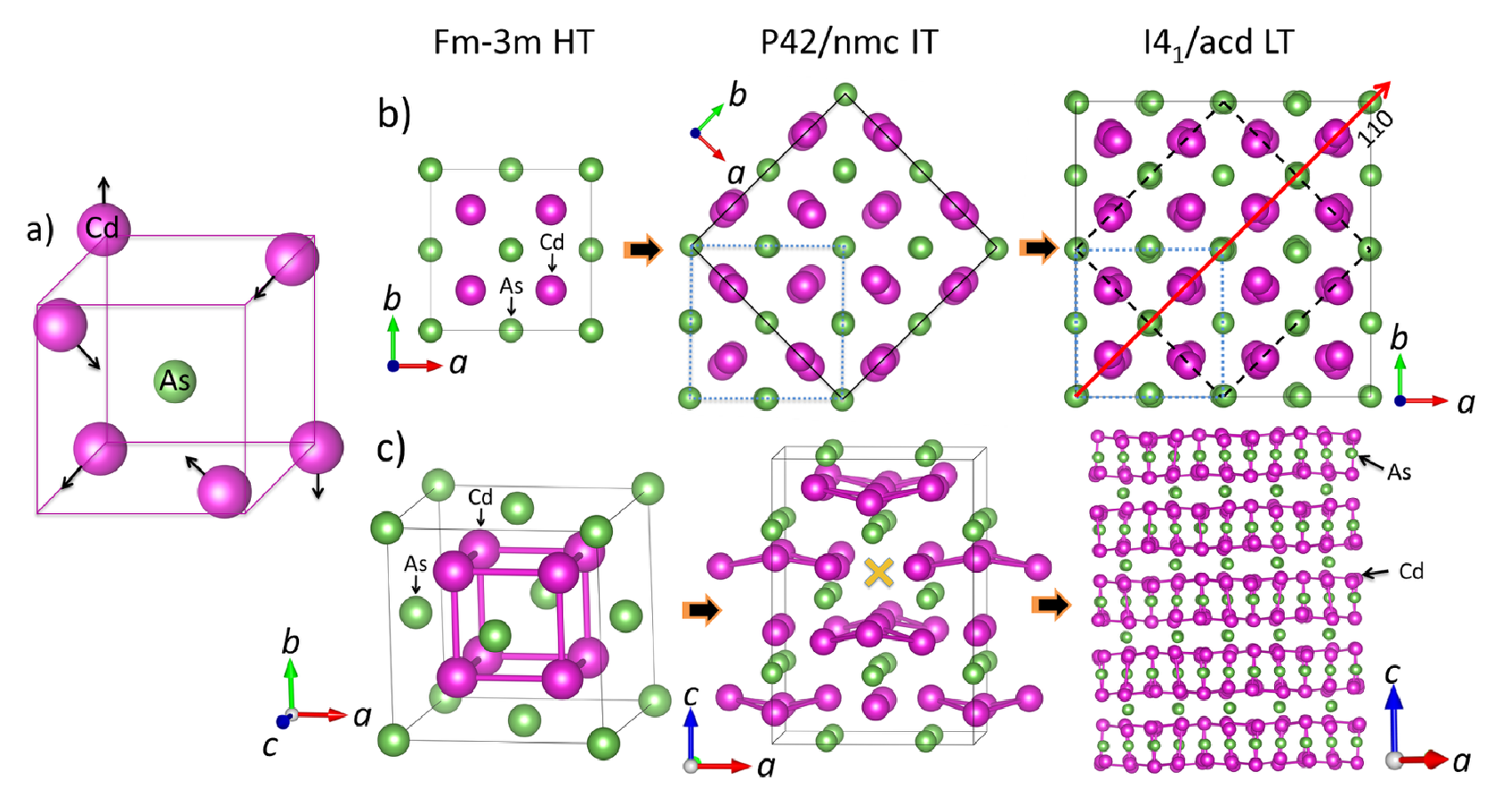}
	\label{Figure_TOC}
\end{figure}

   The structure of Cd$_3$As$_2$, a high mobility semimetal reported to host electrons that act as Dirac particles, is reinvestigated by single crystal X-ray diffraction. It is found to be centrosymmetric rather than noncentrosymmetric as previously reported. It has a distorted superstructure of the antifluorite (M$_2$X) structure type with a tetragonal unit cell of \textit{a} = 12.633(3) and \textit{c} = 25.427(7) \AA~in the centrosymmetric I4$_1$/acd space group. Electronic structure calculations performed using the experimentally determined centrosymmetric structure are similar to those performed with the inversion symmetry absent, but with the important implication that Cd$_3$As$_2$ is a 3D-Dirac semimetal with no spin splitting; there is a four-fold degenerate bulk Dirac point at the Fermi Energy along $\Gamma$-Z in the Brillouin zone. This makes Cd$_3$As$_2$ a 3D electronic analog to graphene.

\end{document}